\documentstyle[12pt]{article}

\def\Pom{{\bf I\!P}}

\def\lsim{\mathrel{\rlap{\lower4pt\hbox{\hskip1pt$\sim$}}
    \raise1pt\hbox{$<$}}}         

\def\gsim{\mathrel{\rlap{\lower4pt\hbox{\hskip1pt$\sim$}}
    \raise1pt\hbox{$>$}}}         

\begin{document}
\begin{center}
\phantom{.}{\large\bf FZJ-IKP-99/04}
\vspace{2.0cm}\\
{\Large \bf Diffractive $S$ and $D$--wave vector mesons in deep inelastic
scattering}
\vspace{2.0cm}\\
{\large  I.P.Ivanov$^{1,2)}$ and N.N. Nikolaev$^{1,3)}$
\vspace{0.5cm}\\}
{\sl
$^{1}$IKP(Theorie), KFA J{\"u}lich, D-52428 J{\"u}lich, Germany \medskip\\
$^{2)}$Novosibirsk University, Novosibirsk, Russia\medskip\\
$^{3}$L. D. Landau Institute for Theoretical Physics, GSP-1,
117940, \\
ul. Kosygina 2, Moscow 117334, Russia}
\vspace{1.0cm}


{\bf Abstract}
\end{center}

We derive  helicity amplitudes for diffractive leptoproduction
of the $S$ and $D$ wave states of vector mesons. We predict a dramatically 
different spin dependence for production of the $S$ and $D$ wave vector
mesons. We find very small $R=\sigma_{L}/\sigma_{T}$ and abnormally 
large higher twist effects in production of longitudinally polarized 
$D$-wave vector mesons. 
\vspace{1.0cm}\\

Diffractive vector meson production
$
\gamma^*+p\to V+p',
$
in deep inelastic scattering (DIS) at small $x={(Q^{2}+m_{V}^{2})}/
{(W^{2}+Q^{2})}$ is a testing ground of ideas on the QCD pomeron
exchange and light-cone wave function (LCWF) of vector mesons
(\cite{KNNZ93,NNZscan,Ginzburg,NNPZ96,NNPZZ98}, for the recent review
see \cite{Review}). (For the kinematics see fig.~1, $Q^{2}=-q^{2}$ 
and $W^{2}=(p+q)^{2}$ are standard DIS variables). The ground state 
vector mesons, $V=\rho^0,\omega^{0},\varphi^{0},J/\Psi,\Upsilon$ are usually
supposed  
to be the $S$-wave spin-triplet $q\bar{q}$ states. However, all the previous
theoretical calculations used the $V\bar{q}q$ vertex $\phi_{V} V_{\mu}
\bar{q}\Gamma_{\mu}q$ with the simplest choice $\Gamma_{\mu}=
\gamma_{\mu}$, which corresponds to a certain mixture of the $S$- and 
$D$-wave states, and any discussion of the impact of the $D$-wave 
admixture in the literature is missing (here $V_{\mu}$ is the vector 
meson polarization vector and $\phi_{V}$ is the vertex function related 
to the vector meson LCWF as specified below.). 

We report here a derivation of helicity amplitudes for diffractive 
production of pure $S$ and $D$-wave $q\bar{q}$ systems for small to 
moderate momentum transfer ${\bf \Delta}$ within the diffraction cone.
Understanding production of $D$-wave states is a topical issue for 
several reasons. First, the $D$-wave admixture may affect predictions
for the ratio $R=\sigma_{L}/\sigma_{T}$, in which there is a persistent 
departures of theory from the experiment. To this end we recall that 
the nonperturbative long-range pion-exchange between light quarks and 
antiquarks \cite{Riska} is a natural source of $S$-$D$ mixing in the 
ground state $\rho^{0}$ and $\omega^{0}$ mesons. Second, different spin 
properties of $S$- and $D$-wave production may facilitate as yet 
unresolved $D$-wave vs. $2S$-wave assignment of the $\rho'(1480)$ 
and $\rho'(1700)$ and of the $\omega'(1420)$ and $\omega'(1600)$ mesons.

In our analysis we rely heavily upon the derivation \cite{KNZ98} of 
amplitudes of the $s$-channel helicity conserving (SCHC) and 
non-conserving (SCHNC) transitions, albeit in slightly different 
notations. We predict a dramatically different spin dependence for 
production of the $S$ and $D$ wave states, especially the  $Q^{2}$ 
dependence of $R=\sigma_{L}/\sigma_{T}$ which derives from the 
anomalously large higher twist effects in the SCHC amplitude for 
production of longitudinally polarized vector mesons. Our technique 
can be readily generalized to higher excited states, $3^{-}$ etc, 
leptoproduction of which is interesting for the fact that they  
cannot be formed in $e^{+}e^{-}$ annihilation. 

A typical leading log${1\over x}$ (LL${1\over x}$) pQCD diagram for vector
meson production is shown in Fig.~1. We use the standard 
Sudakov expansion of all the momenta in the two lightcone vectors 
$$p'=p-q\frac{p^2}{s},\quad q'=q+p'\frac{Q^2}{s}$$ such that $q'^2= p'^2=0$ 
and $s=2p'\cdot q'$, and the two-dimensional transverse component:
$k = z q'+y p'+k_\bot,\quad \kappa=\alpha q'+\beta p'+\kappa_\bot, \quad
\Delta=\gamma p'+\delta q'+ \Delta_\bot\,$ (with the exception of ${\bf r}$
which is a 3-dimensional vector, see below, hereafter ${\bf k},
{\bf \Delta},..$ always stand for 2-dimensional $k_{\perp}, \Delta_{\perp}$ 
etc.). The diffractive helicity  amplitudes take the form
\begin{eqnarray}
A^{S,D}_{\lambda_{V}\lambda_{\gamma}}(x,Q^{2},{\bf \Delta})= 
is {C_{F} N_{c} c_{V}\sqrt{4\pi\alpha_{em}} 
\over 2\pi^{2}}
\int_{0}^{1} {dz\over z(1-z)} \int d^2{\bf k}\psi_{S,D}(z,{\bf k}) 
\nonumber\\
\int {d^{2}{\bf \kappa}
\over
\kappa^{4}}\alpha_{S}({\rm max}\left\{\kappa^{2},
{\bf k}^{2}+\overline{Q}^{2}\right\})
I^{S,D}_{\lambda_{V}\lambda_{\gamma}}(\gamma^{*}\to V)
\left(1+i{\pi \over 2}{\partial \over \partial \log x}\right)
{\cal{F}}(x,{\bf \kappa},{\bf \Delta})
\, ,
\label{eq:8}
\end{eqnarray}
where $\lambda_{V},\lambda_{\gamma}$ stand for helicities, $m$ is the 
quark mass, $C_{F}={N_{c}^{2}-1 \over 2N_{c} }$ is the Casimir operator, 
$N_{c}=3$ is the number of colors, $c_{V}={1\over \sqrt{2}},
{1\over 3\sqrt{2}},{1\over 3},{2\over 3}$ for the $\rho^{0},\omega^{0},
\phi^{0},J/\Psi$ mesons, $\alpha_{em}$ is the fine structure constant, 
$\alpha_{S}$ is the strong coupling and $\overline{Q}^{2}=m^{2} 
+ z(1-z)Q^{2}$ is the relevant hard scale. To the LL${1\over x}$ the 
lower blob is related to the unintegrated 
gluon density matrix ${\cal{F}}(x,{\bf \kappa},{\bf \Delta})$ 
\cite{NNPZZ98,NZsplit,Lipatov}. For small ${\bf \Delta}$ within the 
diffraction cone
\begin{equation}
{\cal{F}}(x,{\bf \kappa},{\bf \Delta})=
{\partial G(x,\kappa^{2})\over \partial \log \kappa^{2}}
\exp(-{1\over 2}
B_{3\Pom}{\bf \Delta}^{2})\,.
\label{eq:9}
\end{equation}  
where  $\partial G/\partial \log \kappa^{2}$ is the conventional
unintegrated gluon structure function and, modulo to a slow Regge growth,
the diffraction cone $B_{3\Pom}\sim$ 6 GeV$^{-2}$ \cite{NNPZZ98}.

In the light--cone formalism \cite{Ter}, one first computes the production 
of an on-mass shell $q\bar{q}$ pair of invariant mass $M$ and total 
momentum $q_{M}$. This amplitude is projected onto the state 
$(q\bar{q})_{J}$ of total angular momentum $J=1$ using the running 
longitudinal and the usual transverse polarization vectors
\begin{eqnarray} 
V_{L}={1\over M}\left(q'+ {{\bf \Delta}^{2}-M^{2} \over s}p'
+\Delta_\bot\right)\,, \quad\quad
V_{T}=V_\bot +{2 ({\bf V}_{\bot}\cdot {\bf \Delta}) \over s}(p'-q'),
\label{eq:4}
\end{eqnarray}
such that $(V_{T}V_{L})=(V_{T}q_{M})=(V_{L}q_{M})=0$. Then the resulting 
upper blob $I(\gamma^{*}\to V)$ is contracted with the radial LCWF of 
the $q\bar{q}$ Fock state of the vector meson, 
\begin{equation}
\psi_{S,D}(z,{\bf k}) = \psi_{S,D}({\bf r}^{2})=
{\phi_{S,D}({\bf r}^2) \over
M^2 - m_V^2}\, .
\label{eq:1}
\end{equation}
Here $r={1\over 2}(k_{2}-k_{1})$, which in the rest frame is  
the relative 3-momentum in the $q\bar{q}$ pair, $r=(0,{\bf r})
=(0,{\bf k},k_{z})$, $r^{2}=-{\bf r}^{2}$, and $$
M^{2}=4(m^{2}+{\bf r}^{2})={m^{2}+{\bf k}^{2}\over z(1-z)}\, .
$$
To conform to this procedure, all the occurrences of the vector meson
mass $m_{V}$ in $I_{\lambda_{V}\lambda_{\gamma}}$ of ref. \cite{KNZ98} 
must be 
replaced by M. 

A useful normalization of the radial LCWF's $\psi_{S,D}({\bf r}^{2})$ is 
provided by
the $V\to e^{+}e^{-}$ decay constant, $\langle 0|J^{em}_{mu}|V\rangle
=fc_{V}\sqrt{4\pi \alpha_{em}}V_{\mu}$:
\begin{eqnarray}
f_{S}={N_{c} \over (2\pi)^{3}}\int d^{3}{\bf r} {8\over 3}(M+m)
\psi_{S}({\bf r}^{2})\, ,
\quad
f_{D}={N_{c} \over (2\pi)^{3}}\int d^{3}{\bf r} {32\over 3}{
{\bf r}^{4}\over M+2m}
\psi_{D}({\bf r}^{2})\, .
\label{eq:3}
\end{eqnarray}

The nice observation is that we need not go again through all the 
calculations of helicity amplitudes. Indeed, the spinor vertices
$\Gamma_{\mu}^{S,D}$ for the pure $S$ and $D$ wave states can be readily
obtained from the simplest $\Gamma_\mu=\gamma_\mu$ used in \cite{KNZ98}.
Following \cite{Ter}, it can be easily shown that 
\begin{equation}
\Gamma^{S}_\mu = \gamma_\mu - {2r_\mu \over M+2m} = 
{\cal S}_{\mu\nu}\gamma_\nu; \quad
{\cal S}_{\mu\nu} = g_{\mu\nu} - {2r_\mu r_\nu \over m (M+2m)}.
\label{eq:6}
\end{equation}
Here we made use of $ r^\mu \gamma_\mu = m$ and $ (q_M\cdot  r) = 0$. 
Once the $S$-wave is constructed, the spinor structure for a $D$-wave 
state can be readily obtained by contracting the $S$-wave vertex with 
$3r_\mu r_\nu + g_{\mu\nu}{\bf r}^2$ with the result
\begin{equation}
\Gamma^{D}_\mu = {\bf r}^2 \gamma_\mu + (M+m)r_\mu = 
{\cal D}_{\mu\nu}\gamma_\nu; \quad
{\cal D}_{\mu\nu} = {\bf r}^2 g_{\mu\nu} + {M+m \over m} r_\mu r_\nu .
\label{eq:7}
\end{equation}
Consequently, the answers for either $S$ or $D$-wave production amplitudes
can be immediately 
obtained from the expressions given in \cite{KNZ98}
by substitutions $  V^*_\mu \to V^*_\nu {\cal S}_{\nu\mu}$, $
V^*_\mu \to V^*_\nu {\cal D}_{\nu\mu} $ for $S$ and $D$-wave states 
respectively.

In terms of diffractive amplitudes ${\bf \Phi}_1$ and $\Phi_2$ defined  
in \cite{KNZ98}, we find for $S$-wave vector mesons (here $T$ stands
for the transverse polarization)  
\begin{eqnarray}
I^S_{0L}= -4 Q M z^2(1-z)^2\Phi_{2}
\left[ 1 + {(1-2z)^2 m \over 2z(1-z) (M+2m)} \right]    
\, ,\nonumber\\
I^S_{T T} =\Biggl\{({\bf V}^{*}{\bf e})[m^{2}\Phi_{2}+
({\bf k}{\bf \Phi}_{1})]
+(1-2z)^{2}({\bf V}^{*}{\bf k})({\bf e}{\bf \Phi}_{1}){M \over M+2m}
\nonumber\\-({\bf e}{\bf k})({\bf V}^{*}{\bf \Phi}_{1})
+{2m \over M+2m}({\bf V}^{*}{\bf k})({\bf e}{\bf k})\Phi_2\Biggr\} 
\, ,
\nonumber\\
I^S_{0T}=-2z(1-z)(2z-1)M({\bf e}{\bf \Phi}_{1})
\left[ 1 + {(1-2z)^2 m \over 2z(1-z) (M+2m)} \right]  
+ {Mm\over M+2m}(2z-1)({\bf e}{\bf k})\Phi_2 
\, ,
\nonumber \\
I^S_{T L}= 2Qz(1-z)(2z-1)({\bf V}^{*}{\bf k})\Phi_{2} 
{M \over M+2m}\,.
\label{eq:10}
\end{eqnarray}
Because the difference between $\Gamma_{\mu}^{S}$ and $\gamma_{\mu}$ is
a relativistic correction, the results for the $S$-wave vector mesons
differ from those found in \cite{KNZ98} only by a 
small relativistic corrections 
$\propto {\bf r}^{2}/M^{2}$. The exceptional case is suppression of
$I^S_{TL}$ by factor $M/(2m+M) \sim 0.5$. 

We skip the twist expansion for $S$-wave amplitides, which
can easily be done following \cite{KNZ98}, and proceed to the
much more interesting case of $D$-wave mesons, for which 
\begin{eqnarray}
I^D_{0L}= -Q M z(1-z)
\left( {\bf k}^2 - {4m \over M}k_z^2 \right)  
\Phi_{2}  
\, ,\nonumber\\
I^D_{TT} =\Biggl\{({\bf V}^{*}{\bf e}){\bf r}^2
[m^{2}\Phi_{2}+({\bf k}{\bf \Phi}_{1})]
+(1-2z)^{2}({\bf r}^2 + m^2 + Mm)
({\bf V}^{*}{\bf k})({\bf e}{\bf \Phi}_{1})
\nonumber\\-{\bf r}^2({\bf e}{\bf k})({\bf V}^{*}{\bf \Phi}_{1})
-m(M+m)({\bf V}^{*}{\bf k})({\bf e}{\bf k})\Phi_2\Biggr\} 
\, ,
\nonumber\\
I^D_{0T}= -{2z-1 \over 2}M
\left\{ ({\bf e}{\bf \Phi}_{1})({\bf k}^2 - {4m \over M} k_z^2)
+ m(M+m)({\bf e}{\bf k})\Phi_2 \right\}
\, ,
\nonumber \\
I^D_{T L}= 2Qz(1-z)(2z-1)({\bf V}^{*}{\bf k})
({\bf r}^2 + m^2 + Mm)\Phi_{2} \,,
\label{eq:11}
\end{eqnarray}
The novel features of these amplitudes are best seen in the twist
expansion in inverse powers of the hard scale 
$\overline{Q}^{2}$. 
As it was noted in \cite{KNZ98}, in all cases but the double
helicity flip the dominant twist amplitudes come from the leading 
log$\overline{Q}^{2}$ (LL$\overline{Q}^{2}$) region of ${\bf k}^{2}\sim 
R_{V}^{-2},{\bf \Delta}^{2} \ll {\bf \kappa}^{2} \ll \overline{Q}^{2}$.
The closer inspection of our $I^{D}_{\lambda_{V}\lambda_{\gamma}}$ shows
that the seemingly leading interference with the dominant $S$-wave component 
in the photon always appears in the quadrupole 
combination $2k_z^2 - {\bf k}^2$. Since the integration over quark loop 
can be cast in form $d^{3}{\bf r}$, such quadrupole combinations 
vanish after angular integration. As a result, the abnormally large 
higher twist contributions $\propto M^{2}/(M^{2}+Q^{2}$ with large
numerical factors come into play and significantly modify 
the $Q^2$ dependence of amplitudes for production of longitudinally
polarized vector mesons:
\begin{eqnarray}
I^D_{0L}=-{Q \over M}\cdot
{32 {\bf r}^4 \over 15 (M^2 + Q^2)^2}\cdot
\left( 1 - 8 {M^2 \over M^2 + Q^2}\right) {\bf \kappa}^2 \, ,\\
I^D_{\pm\pm}=
({\bf V}^*{\bf e})\cdot{32 {\bf r}^4 \over 15 (M^2 + Q^2)^2}
\cdot  \left( 15 + 4 {M^2 \over M^2 + Q^2}\right) {\bf \kappa}^2 \, ,\\
I^D_{\pm L}=
 - {32 {\bf r}^4 \over 15 (M^2 + Q^2)^2}\cdot
{24 Q ({\bf V}^*{\bf \Delta}) \over M^2 + Q^2} {\bf \kappa}^2 \, ,\\
I^{D}_{L\pm}= {32 {\bf r}^4 \over 15 (M^2 + Q^2)^2}\cdot
{8 ({\bf e}{\bf \Delta}) \over M}
 \left( 1 + 3 {M^2 \over M^2 + Q^2}\right)  {\bf \kappa}^2 \,,\\
I^{D}_{\pm\mp}=
 ({\bf V}^*{\bf \Delta})({\bf e}{\bf \Delta}) 
\cdot { 32 {\bf r}^4 \over 15 (M^2 +Q^2)^2}\cdot
\left( 1 - {96 \over 7} {{\bf \kappa}^2 
{\bf r}^2 \over M^2 (M^2 + Q^2)}\right).
\end{eqnarray}
In a close similarity to the $S$-wave case \cite{KNZ98}, the leading 
twist double-helicity flip amplitude is dominated by soft gluon exchange,
the LL$\overline{Q}^{2}$ component is of higher twist.

In order to emphasize striking difference between the $D$-wave and $S$-wave
state amplitudes, we focus on nonrelativistic heavy quarkonia,
where $M^{2}\approx m_{V}^{2}$, although all the qualitative results 
hold for light vector mesons too. For the illustration purposes, we 
evaluated the ratios of helicity amplitudes, 
$\rho_{D/S}=f_{S}A^{D}/f_{D}A^{S}$, for the the harmonic oscillator 
wave functions:
\begin{eqnarray}
\rho_{0L}(D/S)= 
{1\over 5}\left(1-8{m_{V}^{2}\over Q^{2}+m_{V}^{2}}\right)\, ,
\nonumber\\
\rho_{\pm\pm}(D/S)= 
3\left(1+{4\over 15}{m_{V}^{2}\over Q^{2}+m_{V}^{2}}\right)\, ,
\nonumber\\
\rho_{0\pm}(D/S)=-{1\over 5}
(m_{V}a_S)^{2}\left(1+3{m_{V}^{2}\over Q^{2}+m_{V}^{2}}\right)\, ,
\nonumber\\
\rho_{\pm L}(D/S)= 
{3\over 40}(m_{V}a_S)^{4}\, .
\end{eqnarray}
First, $A_{0L}$ changes the sign at  $Q^2 \sim 7 m_{V}^{2}$. The ratio 
$R^{D}=\sigma_{L}/\sigma_{T}$ has thus a non-monotonous $Q^2$ behavior 
and $R^{D} \ll R^{S}$. Furthermore, $R^{D} \lsim 1$ in a broad range of 
$Q^{2} \lsim 225 m_{V}^{2}$. Whereas in heavy quarkonia the $S$--$D$ 
mixing is arguably weak \cite{Novikov}, in light  
$\rho^{0},\omega^{0}$ even a relatively weak $S$-$D$ mixing 
could have a substantial impact on $R$. Second, all the $D$-wave 
amplitides, SCHC and SCHNC alike, with exception of the higher twist 
component of double-helicity flip, are proportional to ${\bf r}^{4}$ 
and, in view of eq.~(\ref{eq:3}), to the decay constant $f_{D}$. In 
contrast to that, in the $S$-wave case the spin-flip amplitudes for
heavy quarkonia are suppressed by nonrelativistic Fermi motion \cite{KNZ98}. 
The relevant suppression parameter is $\sim 1/(a_{S}m_{V})^{2}$, where 
$a_{S}$ is the radius of the $1S$ state. For this reason, for $D$-wave 
states the SCHNC effects are much stronger. For instance, for the 
charmonium $(m_{V}a_S)^{2}\approx 27$, see \cite{Novikov}. 

To summarize, we found dramatically different 
spin properties of diffractive 
leptoproduction of the $S$ and $D$ wave states of vector mesons. We 
predict very small $R^{D}=\sigma_{L}/\sigma_{T}$ and very strong 
breaking of $s$-channel helicity conservation in production of $D$-wave 
states. Higher twist effects in production of longitudinally polarized 
$D$-wave vector mesons are found to be abnormally 
large. Consequently, the distinct 
spin properties of $D$-wave vector mesons in diffractive DIS offer an 
interesting way to discern $S$ and $D$-wave meson states, which are 
indistinguishable at $e^+e^-$ colliders. \\

{\bf Acknowledgments:} The fruitful discussions with B.G.Zakharov 
and V.R.Zoller are gratefully acknowledged. IPI thanks Prof. J.Speth 
for the hospitality at the Institut f. Kernphysik of Forschungszentrum 
J\"ulich. The work of IPI has been partly supported by RFBR.

{\bf Figure caption:}\\

Fig.1: One of the four Feynman diagrams for the vector meson production
$\gamma^{*}p\rightarrow V p'$ via QCD two-gluon pomeron
exchange.

\end{document}